\documentclass[aps,prl,twocolumn,longbibliography,superscriptaddress,floatfix,10pt]{revtex4-2}
\usepackage{hyperref}
\usepackage[utf8]{inputenc}
\usepackage{amssymb,amsfonts,amsmath,dsfont}
\usepackage{graphicx}
\usepackage{textcomp}
\usepackage{gensymb}
\usepackage{pifont}
\usepackage{bm}
\usepackage{color}
\usepackage{accents}
\usepackage{mathtools}
\usepackage{nicefrac}
\usepackage{url}
\usepackage{multirow}
\usepackage[capitalize]{cleveref}
\usepackage{tabularx}
\usepackage{upgreek}
\usepackage{chemformula}
\usepackage{braket}

\hypersetup{colorlinks=true,citecolor=blue,urlcolor=blue,linkcolor=blue}

\usepackage{comment}

\usepackage{mathrsfs}

\newcommand{\bs}[1]{\boldsymbol{#1}}

\makeatletter

\begin{document}

\def \cedenna {Centro  de Nanociencia y Nanotecnología CEDENNA, Avda. Ecuador 3493, Santiago, Chile}
\def \fcfm {Departamento de F\'isica, FCFM, Universidad de Chile, Santiago, Chile.}
\def \usach {Departamento de F\'isica, Universidad de Santiago de Chile }
\def \uta {Departamento de Física, Facultad de Ciencias, Universidad de Tarapacá, Casilla 7-D, Arica, Chile}

\author{Carlos Saji}
\affiliation{\usach}
\author{Mario A. Castro}
\affiliation{\fcfm}
\author{Vagson L. Carvalho-Santos}
\affiliation{Departamento de F\'isica, Universidade Federal de Vi\c cosa, 36570-900, Vi\c cosa, Brazil}
\author{Eduardo Saavedra}
\affiliation{\usach}
\author{Alvaro S. Nunez}
\affiliation{\fcfm}
\author{Roberto E. Troncoso}
\affiliation{\uta}

\date{\today}
\title{Magnetic topological textures in nonorientable surfaces}

\begin{abstract}
Topological magnetic textures confined to two-dimensional (2D) non-orientable manifolds exhibit behaviors absent in planar systems. We investigate bimerons on Möbius surfaces and show that the lack of global orientation alters conservation laws, yielding geometry-dependent topology and dynamics. Micromagnetic simulations reveal that the helical twist and non-orientable geometry reshape the effective topological charge and stabilize chiral configurations imposed by the surface. Under spin-polarized currents, bimerons display unconventional transport: the transverse response is locally reversed or globally suppressed due to charge inversion along the manifold. Moreover, we establish an Aharonov–Bohm effect associated with the magnonic modes of the texture; in particular, the translational Goldstone mode implies that a bimeron on a Möbius strip should exhibit path-dependent quantum interference. These results identify a geometry-driven regime of magnetization dynamics and provide a route to curvature-engineered spintronic functionalities.
\end{abstract}

\maketitle
{{\it Introduction.--} The geometry of the underlying manifold, through its curvature, torsion, and orientability, governs the behavior of physical systems across regimes. On non-orientable surfaces, such as M\"obius strips or Klein bottles, the lack of a global normal direction imposes topological constraints on fields and wave functions, giving rise to geometric phase shifts and modified quantization conditions \cite{Nishiguchi2018,FlourisPRB2022,Wang2022}. In condensed-matter systems, curvature acts as an emergent field that shapes local order and transport, whereas in general relativity it defines the dynamics of spacetime itself. Curvature-induced potentials and torsional effects underlie phenomena ranging from defect formation in liquid crystals \cite{Vitelli,Vitelli-2,dePablo,Napoli} to quantum confinement and topological band structures in low-dimensional materials \cite{daCosta,AokiPRB2001,Ref3}. Extending this framework to non-orientable geometries uncovers new symmetry-protected states and curvature-driven responses, establishing a unified description that links geometry, topology, and material properties, with implications for superconductivity and electronic transport in curved graphene \cite{Ref7,Ref8,Ref9,Vozmediano}.

Curvature and torsion affect magnetization dynamics in nanoscale magnetic systems, establishing the foundation of curvilinear micromagnetism \cite{BookCurve}. Transitioning from planar to three-dimensional geometries introduces curvature as a controllable parameter that links the topology of the magnetization field to the underlying manifold, enabling the stabilization of nontrivial magnetic textures and unconventional responses \cite{Example-1,Example-2,Example-3,Example-4,Pacheco,Experiment-Ribbon}. Curvature modifies exchange-driven anisotropy and Dzyaloshinskii–Moriya interactions \cite{Yershov-Scipost}, while torsion introduces a chiral coupling between the global magnetization and local twist, giving rise to a magnetochiral effect that breaks symmetry and selects a preferred magnetization direction \cite{BookCurve}. 
\begin{figure}[!htb]
\centering
\includegraphics[width=0.9\linewidth]{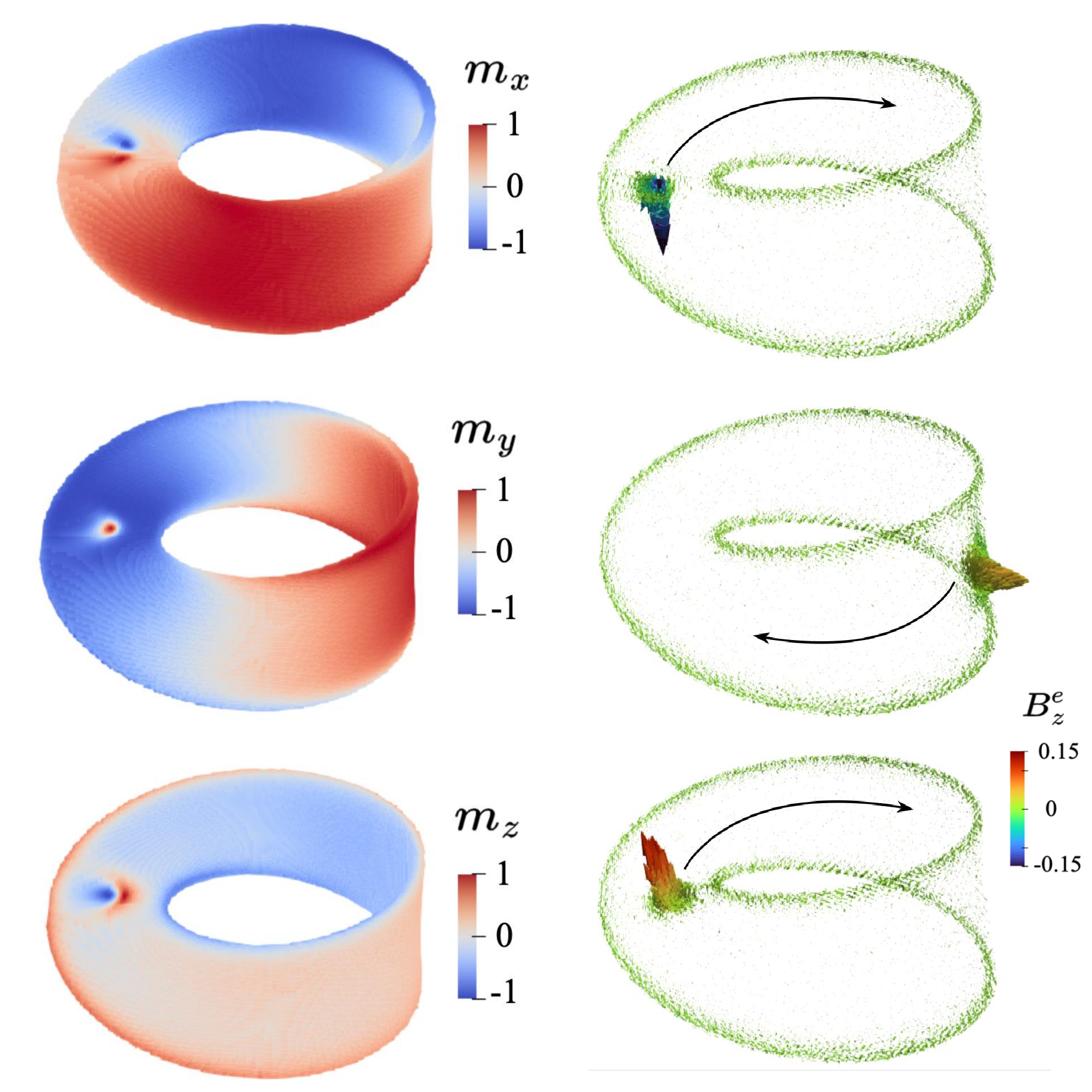}
\caption{Magnetization components of a topological bimeron texture (left panels) stabilized on a M\"obius surface. The corresponding emergent magnetic field ${B}^{e}_n$, locally perpendicular to the surface, is shown in the right panels at successive times during its evolution.}
\label{Fig1-estabilization}
\end{figure}

Twisting further induces magnetoelastic deformations and topological defect formation \cite{Pacheco}, affecting domain-wall dynamics and curvature-driven chirality in Möbius and helical nanostructures \cite{Yershov-2016,Bittencourt-NanoFut,Pylypovsky-PRL}. Non-orientability has been shown to produce nonreciprocal spin-wave propagation under a global geometric twist \cite{dAquino2025}. Despite extensive studies of topological solitons in curved geometries \cite{Wang2019,Galvez2023,Albertino,Yang-SciRep,Garcia,Castro2025}, the influence of torsion and non-orientability on their stability and transport remains largely unexplored.

In this Letter, we study topological magnetic textures confined to nonorientable manifolds, focusing on Möbius geometries (Fig. \ref{Fig1-estabilization}). Surface non-orientability reshapes the definition and conservation of topological charge, stabilizing chiral bimeron states driven by intrinsic helical twist. These textures remain robust across broad parameters and display geometry-dependent Hall effects under spin-polarized currents \cite{Flouris2022}, showing that non-orientable magnetism enables geometry-controlled topological transport. In the semiclassical limit, magnonic modes acquire an Aharonov–Bohm–type geometric phase, rendering the translational Goldstone mode path-dependent and producing geometry-induced interference for a bimeron on a Möbius strip.

Magnetic textures are described by smoothly varying, position-dependent vector fields ${\bf M}({\bf r},t)$, forming spatial magnetization patterns. To capture their geometric and robust topological features on non-orientable manifolds, we evaluate the associated emergent fields and define the corresponding topological charge.
Thus, we construct the double-cover surface $\widetilde{\mathcal{M}}$ which is orientable and periodic \cite{Hatcher}, see Supplemental Material (SM) for details. The double-covering construction untwists the Möbius strip, where two aligned copies cancel the intrinsic reversal, producing an ordinary cylinder. From the emergent magnetic field, $B^{e}_{\mu}=\epsilon_{\mu \nu \rho} \mathbf{m} \cdot\left(\partial_\nu \mathbf{m} \times \partial_\rho \mathbf{m}\right)$, with $\mathbf{m}=\mathbf{M}/M_s$ and $M_s$ being the magnetization saturation, one construct the curvature two-form as $\mathcal{F}= \mathbf{m} \cdot\left(\partial_{x^{1}}\mathbf{m} \times \partial_{x^{2}} \mathbf{m}\right)\ dx^{1}\wedge d x^{2}$, where $x_1$ and $x_2$ are the proper curvilinear coordinates. For a non-orientable manifold, the topological charge is properly defined in the double cover via the projection map $\Pi: \widetilde{\mathcal{M}} \to \mathcal{M}$ and the pullback $\widetilde{\mathcal{F}} = \Pi^* \mathcal{F}$. Thus, the double cover space, $\widetilde{\mathcal{M}}$, corresponds to the $\mathbb{Z}_2$ bundle of frame orientations over $\mathcal{M}$. The non-orientable topological charge can be expressed directly on $\mathcal{M}$ as,
\begin{align}
\mathcal{Q}_{\mathrm{top}}= \frac{1}{8\pi}\int_{\widetilde{\mathcal{M}}} \widetilde{\mathcal{F}},
\label{NOCharge}
\end{align}
where $\widetilde{\mathcal{F}}$ (see left panel in Fig. \ref{Fig1-estabilization}) naturally plays the role of the non-orientable topological charge density. The prefactor $1/8\pi$ accounts for the double covering, ensuring that $\mathcal{Q}_{\mathrm{top}}$ reduces to the usual $1/4\pi$ normalization on orientable surfaces. This formulation properly accounts for the local orientation reversal inherent to non-orientable surfaces and reduces to the standard definition on orientable manifolds. It is valid for smooth magnetization textures whose characteristic length scales exceed the strip width, ensuring the continuum approximation and well-defined emergent fields.}

We now consider the bimeron as an example of a topological soliton in 2D magnetic systems, owing to its relatively simple nucleation and stabilization in chiral magnets \cite{Yang-PRM,Nagase-NatCom,Yang-SciRep,Penthorn-PRL,Stebliy-JAP,Tokura}. Bimeron far field points along the tangential direction ($\mathbf{m}[\rho\rightarrow{0\,(\infty})]=+\mathbf{\mathbf{e}_\varphi}(-\mathbf{e}_\varphi)$), with $\rho$ being the radial distance from the texture’s center. Bimerons and skyrmions share the same topological charge; however, the latter exhibits a distinct far-field magnetization distribution [$\mathbf{m}_{\text{sk}}[\rho\rightarrow{0(\infty})]=+\mathbf{n}\,(-\mathbf{n})$].

\begin{figure*}
\centering
\includegraphics[width=\linewidth]{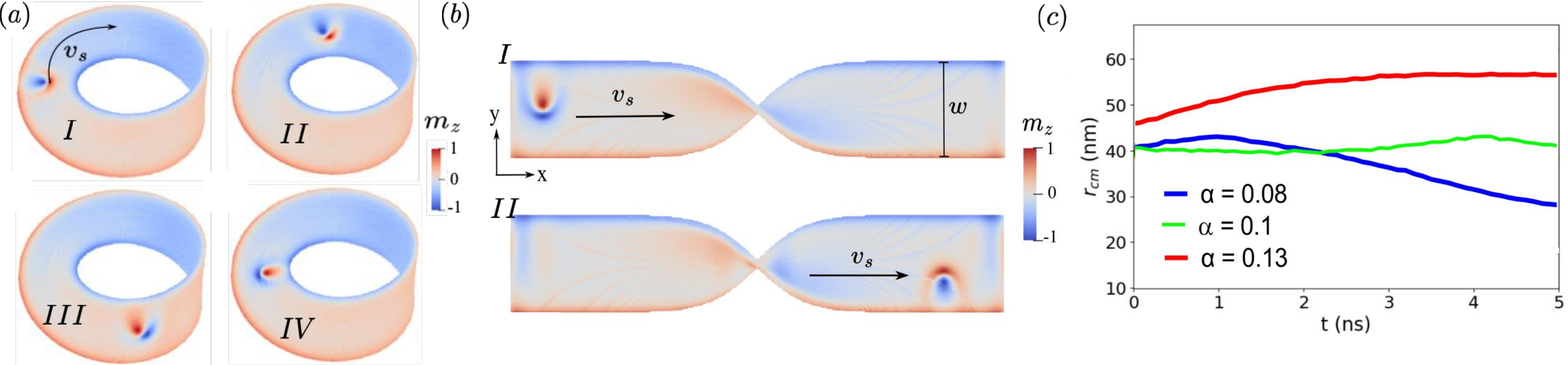}
\caption{Snapshots of the bimeron motion along a M\"obius ring (a) and a straight strip with a torsion (b) under a spin-polarized current with $j = -2 \times 10^{12}$ A/m$^2$, and  $\alpha = \xi = 0.1$. Panels (I)–(IV) show the magnetization profile at increasing times. Due to the non-orientable topology of the M\"obius strip, the bimeron undergoes a continuous reorientation, resulting in a flipped internal structure after one full traversal. (c) Time evolution of the distance between the bimeron’s center of mass and the (initially) outer edge of the Möbius strip under an applied electric current density $j = 2\times 10^{12}$ A/m$^2$, considering that $\xi=0.1$, $P=0.5$, and different values of damping. We observe that, when the skyrmion Hall effect is present ($\alpha\neq \xi$). Depending on whether $\alpha > \xi$ or $\alpha <\xi$, the effective Magnus force initially bends motion towards the inner or the outer edge of the strip, respectively.}
\label{Fig3:bimeron-dynamics}
\end{figure*}

Since the normal vector is ill-defined on a M\"{o}bius surface $\mathcal{M}$, we allow the bimeron to have two states of polarization, both with opposite emergent magnetic vector fields. Accordingly, the collective variable space of a bimeron in this non-orientable surface can be described by a point $\mathbf{X}\in \widetilde{\mathcal{M}}$, with the bimeron core position given by the projection $\mathbf{r}=\pi(\mathbf{X})\in \mathcal{M}$ and its polarization is $\sigma=\pm$ if $\mathbf{X}=\widetilde{\mathbf{r}}_{\pm}$.
Although until now we have focused our discussions on the M\"obius ring, it is worth noticing that the assumptions described here can be extended to a magnetic nanostripe with a localized twist at its center. The surface of this structure, depicted in Fig. \ref{Fig3:bimeron-dynamics}(b), is parametrized by the global coordinates $\mathbf{r}(x_{1},x_{2})=(x_{1},x_{2}\cos(\varphi(x_{1})),x_{2}\sin(\varphi(x_{1})))^{T}$, where we take the function given by  $\varphi(x_{1})={\pi}\tanh(\kappa x_{1})/{2}+{\pi}/{2}$ for $x_{1}\in [-L/2,L/2]$ and $x_{2}\in [-w/2,w/2]$.

The stabilization and current-induced dynamics of a bimeron texture is studied within a micromagnetic approach. The magnetic body is considered as a continuum material whose magnetization vector has the same magnitude but points along different directions throughout it. Firstly, we focus on the bimeron stabilization performing micromagnetic simulations using the GPU-accelerated micromagnetic software package  $\mathrm{Mumax}^{3}$ \cite{Mumax3}, which solves the Landau-Lifshitz-Gilbert equation in the form,
\begin{align}\label{LLEq}
\frac{\partial \mathbf{M}}{\partial t} &=-\gamma \mathbf{M} \times \mathbf{H}_{\mathrm{eff}}+\frac{\alpha}{M_{\mathrm{s}}} \mathbf{M} \times \frac{\partial \mathbf{M}}{\partial t}\,,
\end{align}
where $\gamma$ is the electron gyromagnetic ratio, $\alpha$ is the Gilbert damping constant, and the effective field is given by $\mu_0 {\bm H}_{\text {eff }}=-\partial_{{\bm m}} {E} / M_{\mathrm{s}}$. The magnetic energy is described by the functional $E[{\bs m}]=\int_{V} \mathcal{E}[{\bs m}]\ dV$ where $V$ is the volume of the nanostructure, and the energy density $\mathcal{E}=\mathcal{E}_A+\mathcal{E}_{DM}+\mathcal{E}_d+\mathcal{E}_Z$, includes the contributions of the exchange energy $\mathcal{E}_A=  A(\nabla \mathbf{m})^2$, Dzyaloshinskii-Moriya interaction $\mathcal{E}_{DM}=D \mathbf{m} \cdot(\mathbf{\nabla} \times \mathbf{m})$, magnetostatic interaction $\mathcal{E}_d=-\mu_0 M_s {\bs m} \cdot  \mathbf{H}_{dip}/2$ with $\mathbf{H}_{dip}$ the demagnetizing field, and Zeeman interactions $\mathcal{E}_{Z}=-{\bs m} \cdot \mathbf{B}_{ext}$. 

We adopt material parameters corresponding to FeGe \cite{Yu-Nat,SAAVEDRA2025182703}, which has also been adopted to stabilize bimerons in straight stripes \cite{Castro2025} and nanocylinders \cite{Galvez2023}. The exchange stiffness is set to $A = 8.78$ pJ/m, the Dzyaloshinskii–Moriya interaction constant to $D = 2.9$ mJ/m$^2$, and the saturation magnetization to $M_s = 384\times 10^3$ A/m. The  vacuum permeability is $\mu_0 = 4\pi \times 10^{-7}$ H/m. In addition, an external magnetic field $\mathbf{B}_{\text{ext}}= B_{\phi}\hat{\boldsymbol{\phi}}$ is applied locally tangent to the surface at each point with its magnitude fixed to 600 mT. The spatial discretization used in all simulations is $1\times 1 \times 1$ nm$^{3}$. The non-orientable manifold is characterized by a length $L=160\,\pi$  nm, thickness $t=4$ nm, and width $w=40$ nm. To nucleate the bimeron, we consider an initial magnetization state where magnetization is uniformly magnetized along the local tangential direction ($\mathbf{m}  = \mathbf{e}_\varphi$). Within a small circular region of  20 nm in diameter centered at $(-R,0,0)$, we imprint a Bloch-type bimeron configuration. This initial condition serves to seed the topological texture \cite{Galvez2023}, which is then relaxed until a metastable bimeron state is obtained, as shown in Fig. \ref{Fig1-estabilization}, which depicts the stabilized bimeron (left panel) and its emergent magnetic field (right panel).

Current-driven bimeron motion, is achieved by applying a spin-polarized electric current along the tangential direction of the nanostructure. Accordingly, the magnetization dynamics is described by the Landau–Lifshitz–Gilbert (LLG) equation augmented by the spin-transfer torque terms ($\boldsymbol\Gamma$), given by
\begin{align*}
\boldsymbol{\Gamma}= -\frac{v}{M_{\mathrm{s}}^2} \mathbf{M} \times(\mathbf{M} \times ({\bf j} \cdot \nabla) \mathbf{M})-\frac{\xi v}{M_{\mathrm{s}}} \mathbf{M} \times ({\bf j} \cdot \nabla) \mathbf{M},
\end{align*}
where $v=\frac{P j \mu_B}{e M_{\mathrm{s}}\left(1+\xi^2\right)}$, with $j$, $P$, and $\xi$ stand for the electric current density, the polarization of the spin current, and the non-adiabaticity coefficient, respectively \cite{ZhangLi}. To highlight the geometric effect on the bimeron profile, we set $\alpha = \xi = 0.1$, thereby suppressing the skyrmion Hall effect and enforcing a tangential motion. 

Under the action of a spin-polarized electric current applied along the local tangential direction, the bimeron propagates over the surface along the trajectory shown in Fig. \ref{Fig3:bimeron-dynamics}, which displays snapshots of its evolution on the Möbius strip (a) and on the twisted strip (b). Owing to the non-orientable nature of the geometry, the internal structure of the bimeron undergoes a gradual flip as it traverses the surface. This flip induces a local reversal of the emergent field and thus of the sign of the topological charge density: regions that initially contribute $+1$ locally appear with $-1$ after one full turn. Nevertheless, when the texture is lifted to the orientable double cover, the global topological charge remains $Q_{top} = 1$ throughout the motion, in agreement with the non-orientable definition (see Eq. \ref{NOCharge}). When $\alpha\neq \xi$, the spin-transfer torque generates a finite transverse component of the velocity, producing a skyrmion-Hall–like effect (SHLE). 

In the Thiele picture, the SHLE corresponds to a nonvanishing Magnus force $\mathbf{F}_M\propto \mathbf{G}\times (\mathbf{v}-\mathbf{v}_s)$, where the gyrovector $\mathbf{G}$ is proportional to the local topological charge density. The gyrovector of the bimeron with core position at $\mathbf{R}$ is calculated by the integral of the emergent magnetic field, namely $
\mathbf{G}(\mathbf{R})= \int \mathbf{B}^{e} \ d^{3}{\bf r}$. By considering the planar approximation, where the bimeron radius $r_{0}$ satisfies $r_{0}\ll min(\kappa^{-1}_{1},\kappa^{-1}_{2})$ where $\kappa_{1,2}$ stands for the principal curvature of the surface, it is negligible with respect to the typical length scale of change of the metric. In this case, the gyrovector is evaluated as $\mathbf{G}(\mathbf{X})\approx { 4 \pi h} \mathcal{Q}_{top}\mathbf{n}(\mathbf{X})/\gamma_0$ where $\gamma_0$ is the gyrotropic constant, $\mathbf{n}(\mathbf{X})$ is the normal vector at $\mathbf{X}$, $h$ is the thickness of the surface and $\mathcal{Q}_{top}$ stands for the topological charge of the bimeron, obtained through the double covering space procedure, see Eq. (\ref{NOCharge}).
On a flat strip, the sign of $\mathbf{G}$ fixes the direction of the transverse drift. On a Möbius strip, however, both the local orientation and the emergent field reverse after one turn. Consequently, the effective Magnus force continuously deflects the trajectory toward the same physical edge of the strip, causing the bimeron to propagate along the boundary rather than alternating between opposite sides, as illustrated by the dotted line in Fig. \ref{Fig3:bimeron-dynamics}(c). For the parameters used in this figure, we observe boundary-guided dynamics that depend sensitively on the relative values of $\alpha$ and $\xi$. At $\alpha = 0.13$, the bimeron forms a near-edge channel in which its core propagates parallel to the boundary while maintaining a finite separation from the physical edge, thereby avoiding boundary-induced annihilation. In contrast, at $\alpha = 0.08$ the trajectory approaches the boundary more closely, and the bimeron is ultimately annihilated after a transient regime. This near-edge guiding behavior is consistent with the edge-state picture reported for bimerons in thin magnetic strips \cite{Castro2025}.

Let us consider now the quantum semiclassical dynamics of the bimeron on a M\"obius strip. We consider a magnonic mode localized around the texture, parametrized as $\ket{\Psi_{n}(\boldsymbol{R})}$, with $\boldsymbol{R}$ the position of the bimeron. When the texture completes a loop along the strip, the mode acquires a Berry phase, $\ket{\Psi_{n}(\boldsymbol{R})} \to e^{i\gamma}\ket{\Psi_{n}(\boldsymbol{R})}$, where $\gamma= \oint \boldsymbol{A}(\boldsymbol{R}) \cdot d\boldsymbol{R}$,
and $\boldsymbol{A}(\boldsymbol{R})= \bra{\Psi_{n}(\boldsymbol{R})}i\nabla_{\boldsymbol{R}}\ket{\Psi_{n}(\boldsymbol{R})}$ is the Berry connection.  
The gauge field generated by the texture is determined from the magnon Bogoliubov-de Gennes (BdG) Hamiltonian \cite{DeGennes2018,Zhu2016},
\begin{equation}\label{BdGmagnon}
\mathcal{H}_{\mathrm{BdG}}=\begin{pmatrix}
-(\nabla_{\boldsymbol{r}}-i \boldsymbol{A})^{2}+\mathcal{V} & \mathcal{W} \\ 
\mathcal{W}^{*} & -(\nabla_{\boldsymbol{r}}+i \boldsymbol{A})^{2}+\mathcal{V}
\end{pmatrix}.
\end{equation}
The BdG Hamiltonian is obtained from the Holstein-Primakoff transformation around the texture profile $\boldsymbol{m}(\boldsymbol{r})$, introducing a local orthonormal frame $\left\{\boldsymbol{e}_\Theta, \boldsymbol{e}_\Phi, \boldsymbol{m}\right\}$, defined by the polar and azimuthal vector fields $\boldsymbol{e}_\Theta= \cos \Theta (\cos \Phi\, \boldsymbol{e}_1+\sin \Phi\, \boldsymbol{e}_2)-\sin \Theta\, \boldsymbol{e}_3$  and $\boldsymbol{e}_\Phi=-\sin \Phi\, \boldsymbol{e}_1+\cos \Phi\, \boldsymbol{e}_2 $ \cite{Landeros-DW,Gonzalez2010}. Hence, the spin connection is given by $\boldsymbol{A}  =-\cos\Theta\,(\nabla
\Phi-\boldsymbol{\Omega})- \sin\Theta\,\partial_{\Phi}\boldsymbol{\Gamma}$,  where $\boldsymbol{\Omega}$ denotes the spin connection and $\boldsymbol{\Gamma}=\mathscr{H}\cdot\boldsymbol{\tau}$ with $\mathscr{H}$ the second fundamental form of the surface and $\boldsymbol{\tau}=\cos\Phi\,\mathbf{e}_1+\sin\Phi\,\mathbf{e}_2$ \cite{Gaididei2017,Pylypovsky-PRL}.

In the narrow ribbons approximation, i.e., $t\ll w\ll L$, the spin connection becomes $\boldsymbol{A}(s) \approx \tau_{\mathrm{eff}}$ ($s$ being the length parametrization), with $\tau_{\mathrm{eff}}=\mathcal{C}\pi/L$ the effective torsion for the M\"obius strip \cite{Gaididei2017}, and $\mathcal{C}=\pm 1$ stands for the magnetochirality. Thus, we obtain for the Berry phase $\gamma=\pi \ \mathcal{C} = \pi \ (\mathrm{mod}\ 2\pi)$. Consequently, we establish the existence of an Aharonov–Bohm effect associated with the magnonic modes of the texture. In particular, the above analysis applies to the translational Goldstone mode of the state $\ket{\Psi_{0}(\boldsymbol{R})}$, implying that a bimeron traversing a Möbius strip should exhibit path-dependent quantum interference.


{\it Conclusions.--} We have shown that the non-orientability of Möbius and twisted magnetic nanostructures fundamentally alters the topology and dynamics of magnetic solitons. We focus on bimerons, as stabilizing skyrmions on a Möbius strip requires an additional domain wall imposed by non-orientability \cite{Pylypovsky-PRL}, which would obscure the intrinsic bimeron physics. By formulating the topological charge on the orientable double cover, we demonstrate that bimerons confined to a Möbius manifold undergo a geometry-enforced inversion of their internal structure, recovering their original configuration only after two full traversals of the strip. Micromagnetic simulations confirm that this twist-induced reversal modifies the effective gyrovector, leading to a global suppression—or even reversal—of the skyrmion Hall effect despite sign changes of the local Magnus force along the trajectory. Moreover, we establish an Aharonov–Bohm–type effect associated with bimeron dynamics, which leads to path-dependent quantum interference. Together, these results identify non-orientable geometries as a powerful route to engineering topology-driven transport and interference phenomena in chiral magnetic textures, enabling functionalities inaccessible in planar or orientable curved systems.

{Acknowledgements.--} C.S. thanks the financial support provided by the ANID National Doctoral Scholarship Nº21210450 and Fondecyt Regular 1230747. M.A.C. acknowledges Proyecto ANID Fondecyt Postdoctorado 3240112. E.S. acknowledges support from Dicyt-USACH 042331SD. V.L.C.-S. acknowledges CNPq, Fapemig, INCT/CNPq - Spintr{{\^o}nica e Nanoestruturas Magn\'eticas Avan\c{c}adas (INCT-SpinNanoMag), and Rede Mineira de Nanomagnetismo/FAPEMIG. A.S.N. and R.E.T. acknowledges funding from Fondecyt Regular 1230515 and 1230747, respectively. Funding is acknowledged from the Center for Nanoscience and Nanotechnology, CEDENNA
Project CIA250002.

\bibliography{Bibliography}
\end{document}